\documentstyle[12pt]{article}
\newcommand{\beq}{\begin{equation}}
\newcommand{\eeq}{\end{equation}}
\newcommand{\beqa}{\begin{eqnarray}}
\newcommand{\eeqa}{\end{eqnarray}}
\newcommand{\ba}{\begin{array}}
\newcommand{\ea}{\end{array}}

\begin{document}

\begin{flushright}
Preprint CAMTP/96-4\\
April 1996\\
\end{flushright}

\vskip 0.5 truecm

\begin{center}
\large
{\bf WKB expansion for the angular momentum \\
and the Kepler problem: \\
from the torus quantization to the exact one}\\
\vspace{0.25in}
\normalsize
Marko Robnik$^{(*)}$\footnote{e--mail: robnik@uni-mb.si} and 
Luca Salasnich$^{(*)(+)}$\footnote{e--mail: luca.salasnich@uni-mb.si} \\
\vspace{0.2in}
$^{(*)}$ Center for Applied Mathematics and Theoretical Physics,\\
University of Maribor, Krekova 2, SLO--2000 Maribor, Slovenia\\
\vspace{0.2in}
$^{(+)}$ Dipartimento di Matematica Pura ed Applicata \\
Universit\`a di Padova, Via Belzoni 7, I--35131 Padova, Italy \\
and\\
Istituto Nazionale di Fisica Nucleare, Sezione di Padova,\\
Via Marzolo 8, I--35131 Padova, Italy
\end{center}

\vspace{0.3in}

\normalsize
{\bf Abstract.} We calculate the WKB series for the 
angular momentum and the non--relativistic 3-dim Kepler problem. 
This is the first semiclassical treatment of the angular 
momentum for terms beyond the leading WKB approximation.
We explain why the torus quantization (the leading WKB term)
of the full  problem is exact, even if the individual torus
quantization of the angular momentum and of the radial Kepler
problem separately is not exact. 
\vspace{0.3in}

Submitted to {\bf Journal of Physics A: Mathematical and General}
\normalsize
\vspace{0.1in}
  
\newpage

\section{Introduction}

The semiclassical methods in solving the Schr\"odinger problem
are of extreme importance in understanding the global behaviour
of eigenfunctions and energy spectra, especially as a function
of some external parameter, since usually they are the only
approximation known in the form of an explicit formula. 
\par
The leading semiclassical
approximation is just the first term of a certain $\hbar$-expansion.
The method goes back to the early days of quantum mechanics and
was developed by Bohr and Sommerfeld for one-freedom systems and
separable systems, it was then generalized for integrable 
(but not necessarily separable) systems by Einstein (1917),
which is called EBK or torus quantization. In fact, Einstein's
result was corrected for the phase changes due to caustics by Maslov
(1961; see also Maslov and Fedoriuk 1981), 
but the torus quantization formula thus obtained is still
just a first term in a certain $\hbar$-expansion, whose higher terms
are unknown in systems with more than one degree of freedom. 
Thus recently it has been observed (Prosen and Robnik 1993, 
Graffi, Manfredi and Salasnich 1994) that this leading-order
semiclassical approximations generally fail to predict the
individual energy levels (and the eigenstates) within a vanishing
fraction of the mean energy level spacing. This conclusion is
believed to be correct not only for the torus quantization
of the integrable systems, but also in applying the Gutzwiller trace
formula (Gutzwiller 1990) to general systems, including the completely chaotic
ones, c.f. Gaspard and Alonso (1993). Therefore a systematic study
of the accuracy of semiclassical approximations is very important,
especially in the context of quantum chaos (Casati and Chirikov 1995,
Gutzwiller 1990).
This to end in full generality is an almost impossible task, but
in some  special cases it is possible to work out the quantum corrections
to higher or even all orders (Degli Esposti, Graffi and Herczynski 1991, 
Graffi and Paul 1987, Salasnich and Robnik 1996, Robnik 1984, Narimanov
1995).
On the other hand, in systems with one degree of freedom a systematic WKB
expansion is possible at least in principle, and in a few cases
can be worked out even explicitly to all orders, resulting in
a convergent series whose sum is identical to the exact spectrum
(Dunham 1932, Bender, Olaussen and Wang 1977, Voros 1993, 
Robnik and Salasnich 1996). 
\par
Our goal in the present paper is
to deal systematically with the WKB expansions for the angular
momentum problem and for the Kepler problem. This is important
not only from the point of view of mathematical physics (formal existence
of the systematic series, its convergence properties and the summation),
but also because the Kepler problem is so fundamental in 
physics. To the best of our knowledge a detailed analysis
of this problem has not been undertaken in the literature
so far. 
\par
We shall work out some next to the leading terms for the
Kepler problem and show - under
a conjecture about the higher terms -  that exact result is
obtained after all corrections have been taken into account
and the resulting series has been summed. This is nontrivial,
because we know that the torus quantization of the 3-dim Kepler problem
yields exact result, whereas the individual torus quanization of
the radial and of the angular momentum problems is not exact.
Thus our present work is the first systematic semiclassical
expansion of the angular momentum problem as a pre-requisite
to the full study of the 3-dim Kepler problem.
\par
To define the language and to introduce the notation we
first give the essential formulas of the torus quantization.
The Hamiltonian of the 3-dim Kepler problem is given by
\beq
H={P_r^2\over 2}+{L^2\over 2 r^2} - {\alpha \over r}
\eeq
where
\beq
L^2 = P_{\theta}^2 + {P_{\phi}^2 \over \sin^2{(\theta )}}
\eeq
and
\beq
P_{\phi}=L_z
\eeq
are constants of motion. Of course, the Hamiltonian is a constant 
of motion, whose value is equal to the total energy $E$. 
\par
It is well known that the exact energy spectrum can be obtained 
with the Bohr-Sommerfeld (torus) quantization. 
To perform the torus quantization it is necessary to introduce 
the action variables
\beq
I_{\phi}={1\over 2\pi}\oint P_{\phi} d\phi = P_{\phi},
\eeq
\beq
I_{\theta}={1\over 2\pi}\oint P_{\theta} d\theta =L-I_{\phi},
\eeq
\beq
I_r={1\over 2\pi}\oint P_r dr ={\alpha \over \sqrt{-2E}}-L.
\eeq
The Hamiltonian as a function of the actions reads
\beq
H={-\alpha^2 \over 2 [I_r + I_{\theta} + I_{\phi}]^2},
\eeq
and after the torus quantization
\beq
I_r = (n_r + {1\over 2})\hbar , \;\;\;\; 
I_{\theta}= (n_{\theta} + {1\over 2})\hbar , \;\;\;\;
I_{\phi}= n_{\phi} \hbar ,
\eeq
the energy spectrum is given by
\beq
E_{n_r l} = {- \alpha^2 \over 2 \hbar^2 [n_r + l + 1]^2},
\eeq
where $l=n_{\theta} + n_{\phi}$. 
(Each of the three quantum numbers is a nonnegative integer, and
so is $l$.) This is the exact energy spectrum, 
which can also be obtained by solving the Schr\"odinger equation. 
Note that we have quantized the angular momentum $L=I_r + I_{\theta}$ 
with a semiclassical formula $L=(l+1/2)\hbar$. If we use the exact 
quantization of the angular momentum, i.e. $L=\hbar\sqrt{l(l+1)}$, 
we obtain a wrong formula. How to explain this observation? 
\par
In section 2 we treat the angular momentum problem by calculating the
corrections to the leading torus quantization term, and in 
the section 3 we then proceed with the analysis of the radial
Kepler problem, again by calculating the corrections to the
leading torus quantization term, now using the exact result
for the quantized angular momentum. In section 4 we discuss
the results and draw some general conclusions.

\section{WKB expansion for the angular momentum}

We consider the eigenvalue equation of the angular momentum
\beq
{\hat L}^2 Y(\theta ,\phi ) = \lambda^2 \hbar^2 Y(\theta ,\phi ),
\eeq
where ${\hat L}^2$ is formally given by the equation (2) with 
\beq
{\hat P}_{\theta}^2= -\hbar^2 ({\partial^2 \over \partial \theta^2} 
+ \cot{(\theta )} {\partial \over \partial \theta}) ,
\eeq
\beq
{\hat P}_{\phi}^2= -\hbar^2 {\partial^2 \over \partial \phi^2} .
\eeq
We can write the eigenfunction as
\beq
Y(\theta ,\phi ) = T(\theta ) e^{i n_{\phi} \phi}, 
\eeq
and we obtain
\beq
{\hat P}_{\phi}^2 Y(\theta ,\phi ) = n_{\phi}^2 \hbar^2 Y(\theta ,\phi ) ,
\eeq
and also
\beq
T{''}(\theta ) + \cot{(\theta )} T{'}(\theta ) 
+ (\lambda^2 - {n_{\phi}^2 \over \sin^2{(\theta )} })T(\theta )=0.
\eeq
Notice that $\hbar$ does not appear in this equation anymore.
To perform the WKB expansion we introduce a small parameter $\epsilon$,
which might be thought as proportional to $\hbar$, 
and consider the eigenvalue problem
\beq
\epsilon^2 T{''}(\theta ) + \epsilon^2 \cot{(\theta )} T{'}(\theta ) 
= Q(\theta) T(\theta ),
\eeq
where 
\beq
Q(\theta )= W(\theta ) - \lambda^2 
={n_{\phi}^2 \over \sin^2{(\theta )}}-\lambda^2 .
\eeq
Thus small $\epsilon$ limit is equivalent to the large $n_{\phi}$ and/or 
large $\lambda$ limit. 
The parameter $\epsilon$ helps to organize the WKB series; 
we set $\epsilon =1$ when the calculation is completed. 
First we put
\beq
T(\theta )= \exp{ \{ {1\over \epsilon} S(\theta )\} }, 
\eeq
where $S(\theta )$ is a complex function that satisfies the differential 
equation
\beq
S{'}^2(\theta ) + \epsilon S{''}(\theta ) 
+ \epsilon \cot{(\theta )} S{'}(\theta ) = Q(\theta) .
\eeq
The WKB expansion for the function $S(\theta )$ is given by
\beq
S(\theta ) = \sum_{k=0}^{\infty} \epsilon^k S_k(\theta ), 
\eeq
and by comparing like powers of $\epsilon$ we obtain a recursion 
formula ($n>0$)
\beq
S{'}_0^2 = Q , \;\;\;\;
\sum_{k=0}^n S{'}_k S{'}_{n-k} + S{''}_{n-1} 
+ \cot{(\theta )} S{'}_{n-1} =0.
\eeq
Straightforward calculations give for the first few terms
\beq
S{'}_0= - Q^{1\over 2} ,
\eeq
\beq
S{'}_1 = -{1\over 4} Q{'} Q^{-1} - {1\over 2} \cot{(\theta )} ,
\eeq
\beqa
S{'}_2 & = & -{1\over 32} Q{'}^2Q^{-5/2} 
-{1\over 8} {d\over d\theta} ({ Q{'} Q^{-3/2}}) 
- {1\over 8} \cot^2{(\theta )} Q^{-1/2} \nonumber \\
& - & {1\over 4} ({d\over d\theta} \cot{(\theta )}) Q^{-1/2}.
\eeqa 
The exact quantization of the wave function (18) is given by 
\beq
\oint dS = \sum_{k=0}^{\infty} \oint dS_k = 2 \pi i \; n_{\theta} ,
\eeq
where we have now set $\epsilon =1$. This integral is a complex contour 
integral which encircles the two turning points on the real axis. 
Obviously, it is derived from the requirement of the uniqueness of the 
complex wave function $T$ (Dunham 1932, Bender, Olaussen and Wang 1977). 
\par
The zero order term is given by
\beq
\oint dS_0 = 2 i \int dr \sqrt{\lambda^2 - W(\theta )} 
=2\pi i (\lambda - n_{\phi}),
\eeq
and the first term reads 
\beq
\oint dS_{1} = -{1\over 4} \ln{Q}|_{contour} = - \pi i. 
\eeq
Evaluating $\ln{Q}$ once around the contour gives $4\pi i$ because 
the contour encircles two simple zeros of $Q$. 
\par
All the other odd terms vanish when integrated along the closed 
contour because they are exact differentials (Bender,
Olaussen and Wang 1977). So the quantization condition (25) 
can be written as
\beq
\sum_{k=0}^{\infty} \oint dS_{2k} = 2 \pi i (n_{\theta} 
+{1\over 2})  ,
\eeq 
and thus  it is a sum over even--numbered terms only. 
The next non--zero term is given by
\beqa
\oint dS_2 & = & - i \big[ 
{1\over 12} {\partial^2 \over \partial (\lambda^2)^2} 
\int d\theta {W{'}^2(\theta )\over \sqrt{\lambda^2 - W(\theta )}} 
+ {1\over 2} {\partial \over \partial (\lambda^2)} 
\int d\theta {W{'}(\theta) \cot{(\theta )} \over \sqrt{\lambda^2 - W(\theta )}} 
\nonumber \\
& + & {1\over 4} \int d\theta {\cot^2{(\theta )} \over 
\sqrt{\lambda^2 - W(\theta )}} \big] .
\eeqa
These three integrals give (see the Appendix A)
\beq
\oint dS_2 = {\pi i\over 4 \lambda} ,
\eeq
where, importantly,  the $n_{\phi}$ dependence  drops out now. 
Thus up to the second order in $\epsilon$ the quantization condition reads
\beq
\lambda + {1\over 8 \lambda } = l + {1\over 2} ,
\eeq
where $l=n_{\theta}+n_{\phi}$. The term $1/8\lambda$ is the first quantum 
correction to the the quantization of the angular momentum. 
From this result we can argue ("conjecture by educated guess") 
that the $\epsilon^{2k}$ term in the WKB series is ($k>0$) 
\beq
\oint dS_{2k} = 2\pi i { {1\over 2}\choose k} 2^{-2k} \lambda^{1-2k} ,
\eeq
so that the WKB expansion of the 
angular momentum to all orders is given by
\beq
\sum_{k=0}^{\infty} { {1\over 2}\choose k} 2^{-2k} \lambda^{1-2k} 
=l+{1\over 2} .
\eeq
This is the exact formula for the relationship between $l$ and
$\lambda$, because 
\beq
\sum_{k=0}^{\infty} { {1\over 2}\choose k} 2^{-2k} \lambda^{1-2k} 
={1\over 2}\sqrt{1 + 4 \lambda^2} ,
\eeq
and the equation $\sqrt{1 + 4 \lambda^2} /2 =l+1/2$ can be inverted and 
gives $\lambda =\sqrt{l(l+1)}$. This completes our investigation of the
semiclassical expansion for the angular momentum, where it remains
in general to prove the conjectured formula (32) for $k \geq 2$.

\section{WKB expansion for the radial Kepler problem}

We consider the Schr\"odinger equation for the radial problem
\beq
[-{\hbar^2 \over 2} {d^2 \over dr^2} + V(r)] \psi (r) = E \psi (r)
\eeq
where
\beq
V(r)= {L^2\over 2 r^2} - {\alpha \over r} .
\eeq
We can always write the wave function as
\beq
\psi (r) = \exp{ \{ {i\over \hbar} \sigma (r) \} } ,
\eeq
where the phase $\sigma (r)$ is a complex function that satisfies 
the differential equation
\beq
\sigma{'}^2(r) + ({\hbar \over i}) \sigma{''}(r) = 2 (E - V(r)) .
\eeq
The WKB expansion for the phase is
\beq
\sigma (r) = \sum_{k=0}^{\infty} ({\hbar \over i})^k \sigma_k(r).
\eeq
Substituting (39) into (38) and comparing like powers of $\hbar$ gives 
the recursion relation ($n>0$)
\beq
\sigma{'}_0^2=2(E-V(r)), \;\;\;\; 
\sum_{k=0}^{n} \sigma{'}_k\sigma{'}_{n-k}
+ \sigma{''}_{n-1}= 0 .
\eeq
\par
The quantization condition is obtained by requiring 
the uniqueness of the wave function
\beq
\oint d\sigma = 
\sum_{k=0}^{\infty} ({\hbar \over i})^{k} \oint d\sigma_{k}=
2 \pi n_r \hbar 
\eeq
where $n_r \geq 0$, an integer number, is the radial quantum number. 
\par
The zero order term, which gives the Bohr-Sommerfeld formula (6), 
is given by
\beq
\oint d\sigma_0 = 2 \int dr \sqrt{2 (E - V(r))} ,
\eeq
and the first odd term in the series gives the Maslov corrections
(Maslov index is equal to 2) 
\beq
({\hbar \over i}) \oint d\sigma_{1} = - \pi \hbar. 
\eeq
All the other odd terms vanish when integrated along the closed 
contour because they are exact differentials (Bender, Olaussen and
Wang 1977). So the quantization condition (41) 
can be written
\beq
\sum_{k=0}^{\infty} ({\hbar \over i})^{2k} \oint d\sigma_{2k} = 2 \pi (n_r 
+{1\over 2}) \hbar ,
\eeq 
thus again a sum over even--numbered terms only. 
The next two non--zero terms are (Bender, Olaussen and Wang 1977)
\beq
({\hbar \over i})^{2} \oint d\sigma_2 
= - \hbar^2 {1\over 12} {\partial^2 \over \partial E^2} 
\int dr {V{'}^2(r) \over \sqrt{2 (E - V(r))} } , 
\eeq
\beq
({\hbar \over i})^{4} \oint d\sigma_4 = \hbar^4
[{1\over 240} {\partial^3\over \partial E^3} 
\int dr {V{''}^2(r) \over \sqrt{2 (E - V(r))} } 
- {1\over 576} {\partial^4\over \partial E^4} 
\int dr {V{'}^2(r) V{''}(r) \over \sqrt{2 (E - V(r))} } ] .
\eeq
A straightforward calculation of these terms gives (see the Appendix B)
\beq
({\hbar \over i})^{2} \oint d\sigma_2 = - \hbar^2 {\pi \over 4 L} ,
\eeq
and
\beq
({\hbar \over i})^{4} \oint d\sigma_4 = \hbar^4 {\pi \over 64 L^3} .
\eeq
Up to the fourth order in $\hbar$ the quantization condition reads
\beq
({\alpha \over \sqrt{-2E}} - L) - \hbar^2 {1 \over 8 L} + 
\hbar^4 {1 \over 128 L^3} = (n_r + {1\over 2}) \hbar . 
\eeq
So we have obtained the first two quantum corrections to the torus 
quantization of the radial Kepler problem. Obviously at this point 
of truncating the series we get wrong spectrum if we use the torus 
quantized angular momentum $L=(l+1/2)\hbar$, and this is 
still true if the series is expanded to all orders. However, 
for the anticipated infinite series expansion we shall obtain the 
exact quantized value of the eigenenergies when using the exact 
angular momentum $L^2=l(l+1)\hbar^2$.
To show this we note that 
higher--order corrections quickly increase in complexity but 
each integral gives a polynomial in $E$ with leading term $E^M$, where 
$M$ is the power of the operator ${\partial^M /\partial E^M}$ in front of 
the integral (Barclay 1993). 
Differentiating $M$ times leaves a constant independent of 
$E$. Since this happens in all terms in the series (with $k>0$), 
the WKB corrections to the Bohr-Sommerfeld formula have no E--dependence. 
From this result we can guess the general formula, based on our 
two correcting terms to the torus quantization, namely
\beq
{\alpha \over \sqrt{-2E}} = \hbar \big[(n_r+{1\over 2}) + \lambda + 
\sum_{k=1}^{\infty} {{1\over 2}\choose k} 2^{-2k} \lambda^{1-2k} \big] =
\hbar 
\big[ (n_r+{1\over 2}) + \sum_{k=0}^{\infty} {{1\over 2}\choose k} 2^{-2k} 
\lambda^{1-2k} \big],
\eeq
where $\lambda =L/\hbar$, and so the $\hbar^{2k}$ term in the WKB series 
is ($k>0$)
\beq
({\hbar \over i})^{2k} \oint d\sigma_{2k} =
-2\pi \hbar {{1\over 2}\choose k} 2^{-2k} \lambda^{1-2k} .
\eeq
In conclusion, the energy spectrum of the WKB algorithm to all orders 
is given by
\beq
E^{WKB}_{n_r \lambda} = {- \alpha^2 \over 2 \hbar^2 
[(n_r + {1\over 2}) + \sum_{k=0}^{\infty} {{1\over 2}\choose k} 2^{-2k} 
\lambda^{1-2k} ]^2}.
\eeq
Now, by using the formula (33) of the WKB expansion of the angular momentum, 
we obtain indeed the exact result 
$E^{WKB}_{n_r \lambda}=E_{n_r l}$, as given in equation (9). 
\par
We can summarize the mathematical reason for exactness of the
torus quantization formula (derived in the Introduction) 
for the 3-dim Kepler problem:
Since the problem is separable, the wave functions (for the angular
momentum and for the radial part) multiply and their phases
have the additivity property, and therefore the total
phase written as $\frac{i}{\hbar} (\sigma - i\hbar S)$ must obey
the quantization condition (uniqueness of the wave function).
From the two formulae (32) and (51) one can see that the quantum 
corrections (i.e. terms higher than the torus quantization terms)
do indeed compensate mutually term by term. 

\section{Discussion and conclusions}

In the present paper we offer (to the best of our knowledge) the first
calculation of the higher WKB terms beyond the torus quantization 
leading terms for the angular momentum and the radial Kepler problem.
This analysis explains the curious compensation of the higher order
quantum corrections (of the two separated problems) resulting in the 
exactness of the torus quantization for the entire 3-dim Kepler problem
(see the Introduction). We have no reason to doubt that our conjectured
general formulae (32) and (51) are correct for all $k > 0$, but this still
has to be proved. 
\par
We consider this kind of studies as important in
understanding the accuracy of the semiclassical methods, and much
of the results in this context for 1-dim problems are known, including
some more general families of 1-dim potentials studied
by Barclay (1993) which are characterized by the factorization
property (Infeld and Hull 1957, Green 1965). One important
future project is to analyze a more general class of the 1-dim potentials 
and in particular to extend results to systems with two or more
degrees of freedom, even if they are integrable (but not separable).
Further, it remains as an important project to assess the accuracy
of much more general (although mathematically not yet completely
satisfactory, due to the divergent series expansions) 
methods like the Gutzwiller theory (1967-71, 1990),
applicable to nonintegrable systems, including the chaotic systems
(Gaspard and Alonso 1993).

\section*{Acknowledgements}
LS thanks Professor Sandro Graffi (Universit\`a di Bologna) 
for many enlightening discussions and acknowledges 
the Alps-Adria Rectors Conference Grant of the University of Maribor. 
MR thanks Dr. Evgueni Narimanov and Professor Douglas A. Stone
(Yale University) for stimulating discussions and for communicating
related results. The financial support by the Ministry of Science
and Technology of the Republic of Slovenia is acknowledged with
thanks.
\newpage

\section*{Appendix A}

In this appendix we show how to obtain the formula (30). 
In all integrals of this section the limits of integration are
between the two turning points. 
After substitution $z=\tan{(\theta )}$, we have
\beqa
\int d\theta {W{'}^2(\theta )\over \sqrt{\lambda^2 - W(\theta )}} 
& = & {4 n_{\phi}^4\over \sqrt{\lambda^2 - n_{\phi}^2} }
\int dz {(1+z^2)\over z^6} \sqrt{z^2\over z^2 - \beta } = \nonumber \\
& = & {3 \pi \over 2 n_{\phi}} (\lambda^2 - n_{\phi})^2 
+ 2\pi n_{\phi}(\lambda^2 - n_{\phi}^2) ,
\eeqa
where $\beta = n_{\phi}^2 /(\lambda^2 - n_{\phi}^2 )$, so that 
\beq
{\partial^2 \over \partial (\lambda^2)^2} 
\int d\theta {W{'}^2(\theta )\over \sqrt{\lambda^2 - W(\theta )}} 
= {3 \pi \over n_{\phi}} .
\eeq 
For the other integrals we use the same procedure.
\beq
\int d\theta {W{'}(\theta )\cot{(\theta )}\over \sqrt{\lambda^2 - W(\theta )}} 
= -{2 n_{\phi}^2 \over \sqrt{\lambda^2 - n_{\phi}^2} }
\int dz {1\over z^4} \sqrt{z^2\over z^2 - \beta } 
= - {\pi \over n_{\phi}}(\lambda^2 - n_{\phi}^2) ,
\eeq
from which we obtain
\beq
{\partial \over \partial (\lambda^2)} 
\int d\theta {W{'}(\theta )\cot{(\theta )} 
\over \sqrt{\lambda^2 - W(\theta )}} = - {\pi \over n_{\phi}} .
\eeq 
The last integral gives
\beq
\int d\theta {\cot^2{(\theta )}\over \sqrt{\lambda^2 - W(\theta )}} 
= {1\over \sqrt{\lambda^2 - n_{\phi}^2} }
\int dz {1\over z^2(1+z^2)} \sqrt{z^2\over z^2 - \beta } 
= \pi ({1\over n_{\phi}} - {1\over \lambda}) .
\eeq
In conclusion 
\beq
\oint dS_2 = - i \big[{1\over 12}{3 \pi \over n_{\phi}} + 
{1\over 2}(-{\pi \over n_{\phi}}) 
+ {1\over 4}\pi({1\over n_{\phi}} - {1\over \lambda}) \big] =
{\pi i \over 4 \lambda} .
\eeq

\newpage

\section*{Appendix B}

In this appendix we show how to obtain the formulas (47) and (48). 
In this section again all integrals are taken between the
two turning points. For the first one, after substitution $y=1/r$, we have
\beq
\int dr {V{'}^2(r) \over \sqrt{2 (E - V(r))} }  =   
\int dy {L^4 y^4 - 2L^2 \alpha y^3 + \alpha^2 y^2 
\over L\sqrt{a+ b y - y^2} } , 
\eeq
where $a=2E/L^2$ and $b=2\alpha /L^2$. We observe that 
\beq
I_{2}=\int dy {y^2 \over \sqrt{a+ b y - y^2}}={\pi \over 8}(4a +3b^2),
\eeq
\beq
I_{3}=\int dy {y^3 \over \sqrt{a+ b y - y^2}}={\pi \over 16}(12a +5b^2),
\eeq
\beq
I_{4}=\int dy {y^4 \over \sqrt{a+ b y - y^2}}={\pi \over 128}(48a^2 +128 ab^2 
+35 b^4). 
\eeq
Because we must apply the operator ${\partial^2 /\partial E^2}$ 
and $a=2E/L^2$, the only non--zero contribution stems from the 
integral $I_{4}$ and we obtain 
\beq
{\partial^2 \over \partial E^2} 
\int dr {V{'}^2(r) \over \sqrt{2 (E - V(r))} }  =  {3 \pi \over L} .
\eeq
In conclusion we have
\beq
({\hbar \over i})^2 \oint d\sigma_2 = - \hbar^2 {1\over 12} {3\pi \over L} 
= - \hbar^2 {\pi \over 4 L}.
\eeq
To obtain the formula (48) we proceed in the same way.  
\beq
\int dr {V{''}^2(r) \over \sqrt{2 (E - V(r))} } = 
\int dy {9L^4 y^6 - 12 L^2 \alpha y^5 + 4\alpha^2 y^4\over 
L\sqrt{a+ b y - y^2}},
\eeq
its leading integral is
\beq
I_{6}=\int dy {y^6 \over \sqrt{a+ b y - y^2}}=
{\pi \over 1024}(320 a^3 + 1680 a^2 b^2 + 1260 a b^2 + 231 b^6), 
\eeq
from which we obtain
\beq
{\partial^3\over \partial E^3} 
\int dr {V{''}^2(r) \over \sqrt{2 (E - V(r))} } = {135 \pi \over L^3}. 
\eeq
For the last integral we have
\beq
\int dr {V{'}^2(r) V{''}(r) \over \sqrt{2 (E - V(r))} }  
=\int dy {3L^6 y^8 - 8 L^4 y^7 + 7L^2\alpha^2  y^6 - 2\alpha^3 y^5\over 
L\sqrt{a+ b y - y^2}},
\eeq
its leading integral is
\beqa
I_{8}=\int dy {y^8 \over \sqrt{a+ b y - y^2}} & = &
{\pi \over 32768}(8960 a^4 + 80640 a^3 b^2 + 110880 a^2 b^4 \nonumber \\
& + & 48048 a b^6 +6435 b^8), 
\eeqa
from which we obtain
\beq
{\partial^4\over \partial E^4} 
\int dr {V{'}^2(r) V{''}(r) \over \sqrt{2 (E - V(r))} }  
={315 \pi \over L^3} .
\eeq
In conclusion we have
\beq
({\hbar \over i})^4 \oint d\sigma_4 = \hbar^4 \big[ 
{1\over 240} {135\pi \over L^3} - {1\over 576} {315 \pi \over L^3} \big]
= \hbar^4 {\pi \over 64 L^3}.
\eeq

\newpage

\section*{References} 
\parindent=0. pt
Alvarez G, Graffi S and Silverstone H J 1988 {\it Phys. Rev.} A {\bf 38} 1687 
\\\\
Barclay D T 1993 "Convergent WKB Series", Preprint University of Illinois
at Chicago, UICHEP-TH/93-16.
\\\\
Bender C M, Olaussen K and Wang P S 1977 {\it Phys. Rev. } D {\bf 16} 1740 
\\\\
Casati G and Chirikov B V 1995 {\it Quantum Chaos} (Cambridge: Cambridge 
University Press) 
\\\\
Degli Esposti M, Graffi S and Herczynski J 1991 {\it Ann. Phys.} (N.Y.) 
{\bf 209} 364 
\\\\
Dunham J L 1993 {\it Phys. Rev.} {\bf 41} 713 
\\\\
Einstein A 1917 {\it Verh. Dtsch. Phys. Ges.} {\bf 19} 82
\\\\
Gaspard P and Alonso D 1993 Phys. Rev. A {\bf 47} R3468
\\\\
Graffi S, Manfredi V R and Salasnich L 1994 
{\it Nuovo Cim.} B {\bf 109} 1147 
\\\\
Graffi S and Paul T 1987 {\it Commu. Math. Phys.} {\bf 107} 25 
\\\\
Green H S 1965 {\it Matrix Mechanics} (Groningen: P. Noordhoff)
\\\\
Gutzwiller M C 1967 {\it J. Math. Phys.} {\bf 8} 1979; 
1969 {\it J. Math. Phys.} {\bf 10} 1004; 
1970 {\it J. Math. Phys.} {\bf 11} 1791; 
1971 {\it J. Math. Phys.} {\bf 12} 343 
\\\\
Gutzwiller M C 1990 {\it Chaos in Classical and Quantum Mechanics} 
(New York: Springer) 
\\\\
Infeld L and Hull T H 1951 {\it Rev. Mod. Phys.} {\bf 23} 21
\\\\
Maslov V P 1961 {\it J Comp. Math. and Math. Phys.} {\bf 1} 113--128; 
638--663 (in Russian)
\\\\
Maslov V P and Fedoriuk M V 1981 {\it Semi-Classical Approximations in 
Quantum Mechanics} (Boston: Reidel Publishing Company), and the references 
therein 
\\\\
Narimanov E 1995, private communication
\\\\
Prosen T and Robnik M 1993 {\it J. Phys.} A {\bf 26} L37 
\\\\
Robnik M and Salasnich L 1996 to be submitted
\\\\
Robnik M 1984 {\it J. Phys.} A {\bf 17} 109
\\\\
Salasnich L and Robnik M 1996 "Quantum Corrections to the Semiclassical 
Quantization of a Nonintegrable System", Preprint University of Maribor, 
CAMTP/96-1
\\\\
Voros A 1983 {\it Ann. Inst. H. Poincar\`e} A {\bf 39} 211 

\end{document}